# ZeroTouch Provisioning (ZTP) Model and Infrastructure Components for Multi-provider Cloud Services Provisioning

Yuri Demchenko,
Paola Grosso, Cees de Laat
University of Amsterdam
{y.demchenko, P.Grosso,
C.T.A.M.deLaat}@uva.nl

Sonja Filiposka
Ss. Cyril and Methodius University
in Skopje
sonja.filiposka@finki.ukim.mk
Damir Regvart
CARNET
damir.regvart@carnet.hr

Migiel de Vos
SURFnet
migiel.devos@surfnet.nl
Tasos Karaliotas
GRNET
karaliot@noc.grnet.gr

*Abstract*—This paper presents results of the ongoing development of the Cloud Services Delivery Infrastructure (CSDI) that provides a basis for infrastructure centric cloud services provisioning, operation and management in multi-cloud multi-provider environment defined as a Zero Touch Provisioning, Operation and Management (ZTP/ZTPOM) model. The presented work refers to use cases from data intensive research that require high performance computation resources and large storage volumes that are typically distributed between datacenters often involving multiple cloud providers. Automation for large scale scientific (and industrial) applications should include provisioning of both inter-cloud network infrastructure and intra-cloud application resources. It should provide support for the complete application operation workflow together with the possible application infrastructure and resources changes that can occur during the application lifecycle. The authors investigate existing technologies for automation of the service provisioning and management processes aiming to cross-pollinate best practices from currently disconnected domains such as cloud based applications provisioning and multi-domain high-performance network provisioning. The paper refers to the previous and legacy research by authors, the Open Cloud eXchange (OCX), that has been proposed to address the last mile problem in cloud services delivery to campuses over trans-national backbone networks such as GEANT. OCX will serve as an integral component of the prospective ZTP infrastructure over the GEANT network. Another important component, the Marketplace, is defined for providing cloud services and applications discovery (in generally intercloud environment) and may also support additional services such as services composition and trust brokering for establishing customer-provider federations.

I. INTRODUCTION

Modern research is becoming more and more data intensive and requires using high-performance computing and large volume storage. Cloud Computing [1, 2] and Big Data technologies [3, 4] provide necessary computing and data processing platform for research and industry data intensive and data driven applications. As an example, bioinformatics is dealing with the genome sequencing which is compute intensive and often requires using distributed data sets and computing resources from multiple data centers or cloud providers. The researchers and research institutions are looking for a possibility to automate scientific applications deployment and management as much as it is possible so that they can focus on their main research work/tasks and use complex applications and infrastructure at a "fingertip".

Industry and businesses today, after widely adopting cloud computing, and thus benefiting with reduced capex; are looking to decease their high operational expenses (opex) entailed by the continuously changing applications and business processes in agile data driven companies. They look for applications deployment and operations processes automation.

The growing demand for automation of applications provisioning acts as a motivator for development of a new concept of applications provisioning, operation and management in clouds defined as Zero Touch Provisioning, Operations and Management (ZTPOM). This trend is also recognised by TMForum that have launched the ZOOM (Zero-touch Orchestration, Operations and Management) program in order to develop best practices and standards for a new generation of service provider support systems that will deliver high business agility and rapid new service development [5]. We distinguish ZTPOM from ZOOM as it is more focused on provisioning and operation typical for infrastructure services provisioned on demand.

We envision ZTPOM as a necessary and enabling element of the general (Inter-)Cloud Services Delivery Infrastructure (CSDI) that implements the major architectural patterns of the Intercloud Architecture Framework (ICAF) and the Intercloud Federation Framework (ICFF) proposed in earlier authors' works [6, 7]. The proposed ZTPOM concept and architecture are a continuation of an ongoing research and development within the GEANT project that previously proposed the Open Cloud eXchange (OCX) [8]. OCX is envisioned as an overlay on top of GEANT network infrastructure that address cloud services delivery from multiple Cloud Services Providers (CSPs) to GEANT's user communities encompassing the European research and academic organisations. In fact, OCX solves the "last mile" problem in cloud services delivery, using a community (or corporate) network interconnecting multiple institutions or virtual distributed project oriented research teams. Effective OCX and CSDI operation requires maximum automation of the underlying and interconnected network provisioning and in this way motivates for ZTP oriented services. At the same time, OCX is an important enabling element of ZTPOM in intercloud multi-provider environment. Initial research on this topic is presented in the previous authors' paper [10] that was primarily focused on the network ZTP technologies, and provided initial research on the Zero Touch Network (ZTN) model, requirements and infrastructure components for cloud based applications. The proposed ZTPOM is intended to benefit from the best practices in ZT network provisioning, cloud deployment and management automation.

The remainder of the paper is organized as follows. Section II describes general use cases for data intensive applications that motivate the need for ZTP automation. Section III provides overview of existing solutions and best practices in Zero Touch network automation, cloud automation and solutions. Section IV proposes the ZTPOM architecture and discusses infrastructure components. Section V takes closer look at popular cloud automation tools and provides suggestion about extension of the SlipStream cloud automation platform to make it suitable for ZTPOM. Section VI describes the recent OCX and SDN/ZTN demo at SuperComputing Exhibition 2015. The paper concludes with remarks on future development in section VII.

II. USE CASES FOR DATA INTENSIVE APPLICATIONS

*A. GEANT: Delivering high-performance services to campus users*

The GEANT network is dedicated to deliver high performance cloud based services to university users. However, bringing clouds closer to the user is just one step in currently growing demand for complex distributed applications comprising of two major trends on the cloud based data intensive services scene: (1) the rising popularity of distributed inter-cloud services, going hand in hand with (2) the necessity for an intelligent agile network infrastructure between the providers and the client that will be able to respond to the cloud nature, i.e. mirror the on-demand, scalable and flexible nature of the federated cloud services.

These trends create both the challenge and opportunity for GEANT and NRENs to provide easy to use network infrastructure services environment for their communities: the research and education (R&E) institutions. Data driven and data intensive research employ big data that needs to be processed in a fast, often streaming, way requiring high performance computing power, but also fast, reliable network that will move the data from one place to another depending on the data location or application workflow. Unlike in the in-house HPC era, today most of the storage and processing power are distributed among multiple CSPs and private datacenters/clouds [9]. Thus, GEANT and the NRENs need to provide guaranteed high QoS network connectivity for advanced cloud services access and delivery while still enabling other features such as security and federated access control.

However, the task is not straightforward because of the high degree of heterogeneous actors that come into play. The newest and most popular cloud applications are becoming increasingly complex especially being made up of multi-cloud heterogeneous environments that involve multiple stakeholders (including cloud, network, application and data providers). The complex workflow of these applications involve a number of steps that are executed at different CSPs and in order to achieve high performances for the complete application workflow, all stakeholders must be interconnected using an advanced inter-cloud network that will provide the means for the seamless big data flow from one step to the other.

The work done in the Service Activity 7 within the GEANT4 project has provided the first major step of bringing closer the cloud services and CSP specific features to the R&E community. Based on the developed cloud catalogue, the users can now browse the listed CSPs and search for specific requirements for their envisioned use of the services provided. In order to increase the value of the cloud catalogue, and further improve the cloud service usage, the next step towards the defined goals is to enable automation for the cloud service users. One part of this automation can be achieved using the now existing cloud automation methods, but these must also be employed for the underlying network infrastructure as well so that the complete cloud service delivery infrastructure can be united. Seen from a user centric point of view, this strategy aims for a ZTP approach that will integrate all infrastructure components and offer highly automated solution to the end user which is especially important for data intensive applications.

*B. Bioinformatics*

Bioinformatics represents one of the most demanding use cases for both high-performance computational infrastructure provisioning and applications deployment automation. Bioinformatics generates huge amount of data produced by multiple research teams from the DNA sequencing. Decreasing prices for DNA sequencing that in a single case produces terabytes of information already cause problems for effective data management. In many cases bioinformatics research requires using either few specialised geographically distributed data sets or use specialised applications that are available at one or another research center. So, applications deployment automation and movement of large data volumes are the key problems that the research infrastructures need to solve to increase effectiveness of bioinformatics research. While the CYCLONE project (http://www.cyclone-project.eu/) [11] attempts to address bioinformatics applications automation, the dedicated network infrastructure provisioning will require corresponding ZTP functionality from the GEANT network.

*C. Video processing*

Another example for a highly intensive big data streaming application is real-time, or near real-time, video processing on the fly in the cloud. Video processing becomes a highly data intensive application when working with high resolution content (4K or 8 K) that is captured with a specialized camera in real time and then needs to be preprocessed on the fly before broadcasted to the viewing equipment.

The preprocessing is carried out using an imaging software on a high processing machine that can be spawned in the cloud while the source and the receiver can be on virtually any location within GEANT. The same problem can occur even in non real-time scenarios when the video source can be stored on one location, while the received transformed video is needed by a R&E institution on a different location. The video size and type are such that the required network bandwidth is far from an average best effort Internet consumption and thus require specialized reserved resources. In the real-time case these need to be augmented with requirements for maximum latency and jitter as well.

Furthermore, the preprocessing can include multiple specialized transformations that can be performed on different machines provided by different CSPs. In this case the scenario becomes even more complex since it turns into a multi-cloud application where, depending on the ordering of the transformations, the application workflow defines the path that the real-time big data stream will need to take from the source via the multiple transformation steps in possibly different CSPs up to the receiving end.

The CSDI integrated in the GEANT network is of great importance in this augmented use case scenario since it needs to provide the flexibility to change the big data stream path according to the multi-cloud application workflow. In the most complex scenario the workflow will be defined by the end-user, thus changeable in real time, which means that the only way to enable a flexible CSDI that will respond to this challenge is to extend it with ZTP.

III. OVERVIEW: AUTOMATED PROVISIONING OF INFRASTRUCTURE SERVICES

*A. Network automation*

The ZTP concept is not new in networking and has been employed to provide different levels of automation in the part of network management over time. Recent advances in ZTP have enabled employing concepts from virtual servers fast provisioning to new network devices automated remote provisioning. In this case the concept of plug and play is transferred to the network domain, so that new networking devices are using the bootstrapping principle in order to obtain their global and specific configuration from a given server after they join the network for the first time.

This enables further improvements to the main ZTP idea by enabling the end user to partially influence the final configuration of the network device, which, combined with a fully orchestrated approach that can be achieved on the level of a network domain, can enable agile end to end service delivery under required network constraints. Examples that show this approach in action are the Glue networks engine [12] and the Aruba network orchestration [13, 14]. However, all existing solutions are valid for only one domain, while the extension to the GEANT's multi-domain environment is not straightforward.

*B. Infrastructure centric cloud automation*

Motivated by growing number of use cases and applications that require integrating resources from multiple cloud providers of different type (I/P/SaaS) the traditional telecom providers and network vendors are coming with their infrastructure centric cloud services automations. The examples are Cisco Intelligent Automation for Cloud (IAC) [15] and Equinix Cloud Exchange [16].

*1) Cisco IAC*

Cisco Intelligent Automation for Cloud aims for delivering different service types across mixed environments in a self-service manner [15]. It is a cloud management platform that offers a self-service portal with service catalogue, network services automation using out of the box templates for different network topologies and applications, tenant segregation within shared container environments, provisioning and management that allows extension to additional hypervisors and cloud providers, and easy addition of cloud accelerators via specialized compact solution sockets and API.

*2) Equinix Cloud Exchange and Cisco ICA*

The Equinix Cloud Exchange [16] is a solution offering secure, high performance, reliable connectivity to a group of cloud service providers. By interconnecting with a number of CSPs it provides its customers on-demand private virtual connections for direct access to cloud providers that are either co-located in a Equinix data center or connected via a single port connection through a network service provider.

The self-provisioning of the cloud exchange service is enabled by using Cisco Network Services Orchestrator (NSO) that provides near zero-touch provisioning to customers through automation. It is based on Tail-f's Network Control System [17] product acquisitioned by Cisco, that models services and network in a standardized high level language (YANG) which increases vendor independence. This is an SDN based solution that supports the entire service life-cycle.

Together these two create the Equinix Programmable Network that enables the automated network and service provisioning for cloud providers and their enterprise customers.

IV. ZTPOM ARCHITECTURE AND INFRASTRUCTURE COMPONENTS

Implementing and operating ZTPOM services in a distributed multi-provider and multi-domain environment will require dedicated infrastructure that supports distributed control over the large scale services and resources deployment. In many cases it requires complex workflow execution that defines both the sequence of resource deployment, activation and adjustment to local environments (that will depend on the current state of the provider resources) and requires resources interaction.

While developing a consistent model for ZTP enabled complex multi-cloud/intercloud applications over GEANT and NRENs network infrastructure, we plan to reuse cloud experience gained with a number of large scale services deployment and extend it with the ZTP enabled network infrastructure provisioning supported by Software Defined Networks (SDN) [18] and Network Functions Virtualisation (NFV) [19] at all levels of the network infrastructure.

Figure 1 illustrates the general ZTPOM architecture and its main infrastructure components. In general, the ZTPOM

infrastructure includes a ZTPOM Server (or Engine) for each application domain (can also be coordinated by the central ZTPOM server) and ZTP clients. The server holds the full information about application or infrastructure to be provisioned; this may include application topology, configuration of all elements, and may also contain application VM images (although the latter may be outsourced to application repository). The server initiates clients to run the application or perform infrastructure provisioning locally on the hosting cloud. Clients can be deployed as a part of the initial Virtual Private Cloud (VPC) deployment or installed as a part of application VMs/infrastructure. Within a typical ZTP scenario, the ZTP client may need to discover the ZTP server and download necessary information and images before staring the deployment process or workflow (WF) based on the application blueprint or recipe. It is anticipated that the ZTP process may need to adjust the provided general recipe according to local VPC environment and cloud platform. The full ZTPOM process will also include application operation monitoring and may trigger application re-configuration.

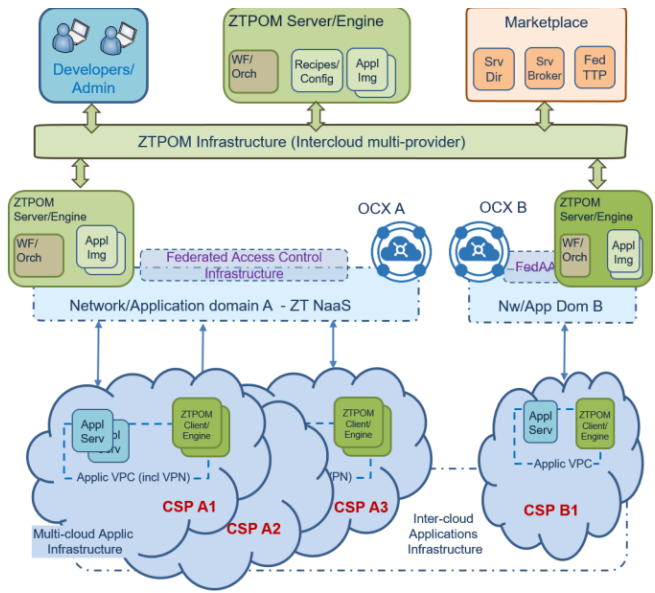

Figure 1. ZTPOM architecture and infrastructure components.

The figure also includes other ZTPOM infrastructure components that are required for applications provisioning and operation in intercloud multi-provider environment such as: intercloud network infrastructure (that can be logically overlaid over the public Internet or dedicated provisioned over corporate network or telecom provider network), OCX, and the Marketplace as a third party service that enables inter-provider services discovery, brokering and trust establishment.

### A. OCX and Zero-touch provisioning ecosystem

Network connectivity provisioning in multi-cloud environment will require dedicated infrastructure components with intra-domain and inter-domain ZTPOM configuration servers, in order to enable automated service provisioning over an already established multi-domain/multi-NREN network infrastructure. This functionality is provided by OCX (as a part of the GEANT connectivity on demand service) that provides connectivity and network information exchange point for cloud services delivery from cloud data centers to user locations in NRENs and campuses. The OCX concept and architecture is validated in multiple demonstrations [8, 20, 21] and its functionality can be directly integrated into the ZTPOM infrastructure.

### B. Marketplace

Marketplace is an important component in ZTPOM as a third party service that allows inter-provider information exchange and brokering (to overcome CSP services model limited to their own infrastructure). It is anticipated that each cloud provider may run own proprietary service similar to marketplace to offer preconfigured VM images or applications, however, in the case of multi-provider applications, separate Marketplace services need to be created to allow bridging/brokering different provider administrative domains.

Typical Marketplace functionality includes: cloud services directory where the connected CSPs publish their services (including SLA, API and X.509 certificates), cloud services brokering, API to "connectivity as a service" service that can be provided by OCX and ZTPOM, and federation provider that provides trusted repository of customer and provider X.509 Certificates with corresponding trust establishment protocol (FedTTP).

For assuring a successful implementation it is important that the marketplace is technology agnostic and flexible enough to cope with most of the requirements from the end-users and CSPs, otherwise they would prefer direct peering to aggregate and provide cloud services with their chosen CSPs.

The marketplace must also include a ZTP API so that the selected user applications/services can be directly provisioned via ZTP infrastructure.

## V. SELECTING ZTPOM PLATFORM AND TOOLS

### A. Cloud Automation Tools

The cloud based applications can be deployed using some of the cloud automation tools like Chef, Puppet or Ansible [22], where, using recipes or cookbooks, the application developer can describe machines configuration in a declarative language, bring them to a desired state, and keep them there through automation. The complete application topology and components interrelationship can be described using a language like OASIS TOSCA (Topology and Orchestration Specification for Cloud Applications) [23], wherein the workflow that invokes different cloud based services is provided.

However, with the current cloud automation tools the problem of provisioning inter-cloud network connectivity remains unsolved, their network configuration capabilities allow only intra-cloud network configuration for one cloud platform or VPC. The solution is seen with adding SDN based network provisioning capability to the proposed ZTPOM.

A good example of fusion between cloud originated technologies and SDN is the recent development of the Network Automation and Programmability Abstraction Layer with Multivendor support (NAPALM) system [24] that implements a common set of functions to interact with different network Operating Systems using a unified API. NAPALM supports several methods to connect to the devices, to manipulate configuration or to retrieve data and uses Ansible [25] to configure network devices as programmable devices. Ansible has benefits against other tools for network deployment and management as it does not require a node agent and runs all process over SSH which simplifies its use for configuring network devices from multiple vendors.

The proposed research in cooperation between GEANT4 and CYCLONE project is developing a new platform for ZTPOM based on SlipStream that intends to solve multi-cloud applications provisioning and management.

### B. SlipStream

SlipStream [26] is an open source cloud application manager. It is used within the CYCLONE project [11] to deploy cloud applications onto one or more IaaS cloud infrastructures and to manage the cloud resources allocated to the running cloud applications. The core SlipStream component is the deployment engine that is augmented with "App Store", "Cloud Service Catalog", dashboard, and monitoring to provide a complete PaaS environment for applications development and operation also referred to as DevOps. SlipStream supports most public IaaS clouds and related API, in particular Amazon Web Services (AWS) and Microsoft Azure cloud. SlipStream also natively supports community research cloud StratusLab and general OpenStack based clouds. To achieve interoperability in the heterogeneous multi-cloud environment, SlipStream is gradually moving from its own RESTful API to the standardized CIMI API [27].

SlipStream development includes integration of inter-cloud network infrastructure provisioning functionality and federated identity and access control infrastructure services such as eduGAIN [28]. This makes SlipStream a platform of choice for implementing ZTPOM infrastructure and engine in particular.

### VI. SC15 DEMO AND TESTBED

A use case scenario based on a multi-cloud application for real-time UHD video editing clearly demonstrates the power of NFV and SDN as the underlying technologies for ZTPOM. The demonstration given at SC15 focused on using NFV combined with Service Function Chaining (SFC) [29] over OCX's programmable network (SDN). NFV was used to modify functions performed over the active traffic, while full transparency and real-time agility was provided using a SDN programmable network.

In the SC15 demo, the gOCX programmable instance directs the traffic flow between the several CSPs, each providing specific video editing function according to the SFC workflow that is defined directly by the end-user in real time. The SFC implementation was based on VLAN stitching and MAC rewriting in order to make sure that different paths and functions are kept separate. The newly proposed protocol for identifying service function paths in the network called Network Services Headers (NSH), should increase the possibilities of SFC even further enabling richer and more complex scenarios.

The achieved high quality of experience (QoE) with a near real-time response in the demo presents the pivotal performances one can expect from a multi-cloud application that runs on a cloud aware programmable network. The lessons learnt from this demo are to be used for the detailed definition of ZTPOM and OCX. The demo has managed to set the grounds for establishing a testbed for multi-cloud high performance applications that rely on a highly agile programmable network that can instantly respond to the requirements defined by the application workflow.

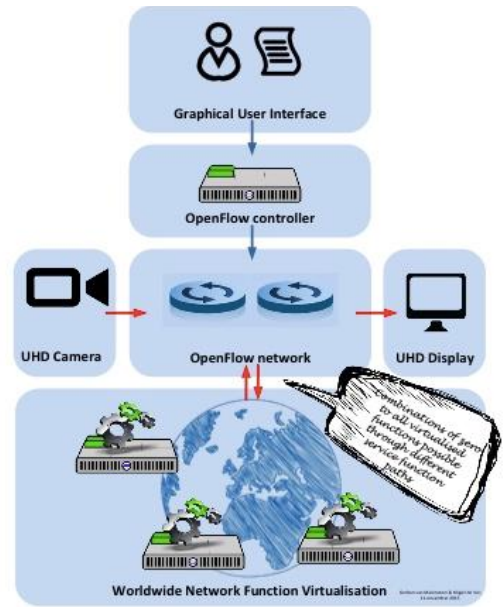

Figure 2. SC15 demonstration scenario

### VII. FUTURE DEVELOPMENT

The proposed ZTPOM over the OCX infrastructure creates a complete ecosystem that fosters the development of new multi-cloud services and application that can directly interact with the underlying cloud aware agile programmable network. Using the NFV paradigm, the OCX marketplace can become a pool of virtualized resources that can be used for defining composite complex applications in a modular way by the means of SFC. In this way the creation of custom multi-cloud solutions used by research community can be pushed towards the zero touch experience for the end-users that can now focus on what they need opposed to how to do it. In this sense, the self-service portal for end-users is envisioned to become a blackboard where using existing standard and/or integrated cloud services, the end user can create a custom workflow that will automatically interact with the agile programmable network infrastructure in order to ensure high performance delivery of targeted data flows whenever necessary. The self-service portal is seen by TMF ZOOM architecture as a part of the next generation operations support and management system.

The further development of the ZTPOM concept will require extensive research in multi-domain SDN with elements of Self-Organized Networks (SON) [30] and efforts in enhancing the current capabilities of NetOps and DevOps that need to become more flexible with an increased number of options for providing capabilities to define the application workflow and automatically provision the network accordingly.

ACKNOWLEDGEMENT

The research leading to these results has received funding from the European Community's Horizon2020 Programme under Grant Agreement No. 691567 (GÉANT 4, Phase 1). Research and developments related to cloud deployment automation are partly supported by the Horizon2020 project CYCLONE (Grant Agreement No. 644925).